\documentclass[10pt,twocolumn,twoside]{IEEEtran} 
\newlength{\pic}
\newlength{\bigPic}
\usepackage{graphicx}
\usepackage{amsthm}
\usepackage{amssymb}
\usepackage{pgf,pgfarrows,pgfnodes}
\usepackage{tikz}
\usepackage{algorithm, algorithmic}
\usepackage{setspace}
\usepackage{lpic}
\usetikzlibrary{arrows,decorations.pathmorphing,backgrounds,positioning,fit,calc} 
\newtheorem{lemma}{Lemma}
\newtheorem{theorem}{Theorem}
\newtheorem{corollary}{Corollary}
\newtheorem{proposition}{Proposition}

\theoremstyle{definition}
\newtheorem{definition}{Definition}
\newtheorem{example}{Example}
\theoremstyle{remark}

\theoremstyle{plain}

\begin{document}

\title{Coding for Cryptographic Security Enhancement using Stopping Sets}

\author{*Willie~K.~Harrison,~\IEEEmembership{Student~Member,~IEEE,}
        Jo\~{a}o~Almeida,~\IEEEmembership{Student~Member,~IEEE,}
        Steven~W.~McLaughlin,~\IEEEmembership{Fellow,~IEEE,}
        and~Jo\~{a}o~Barros,~\IEEEmembership{Member,~IEEE}
\ifCLASSOPTIONpeerreview
\thanks{W. K. Harrison: Mailing address: Georgia Institute of Technology, Atlanta, GA 30332 USA. Affiliation: Same as S. W. McLaughlin. Phone: (+1)435-512-0294. Fax: (+1)404-894-7883. E-mail: harrison.willie@gatech.edu.}
\thanks{S. W. McLaughlin: Mailing address: Georgia Institute of Technology, A. French Building, Suite 111 Atlanta, GA 30332-0740 USA. Affiliation: School of Electrical and Computer Engineering, Georgia Institute of Technology. Phone: (+1)404-385-3383. Fax: (+1)404-385-6690. E-mail: steven.mclaughlin@provost.gatech.edu.}
\thanks{J. Almeida: Mailing address: Instituto de Telecomunica\c{c}\~{o}es (IT-Porto), FEUP, Departamento de Engenharia Electrot\'{e}cnica e de Computadores, Rua do Dr. Roberto Frias, s/n, 4200-465 Porto, Portugal, Visitors: Building I, Office I322. Affiliation: Instituto de Telecomunica\c{c}\~{o}es, Departamento de Engenharia Electrot\'{e}cnica e de Computadores, Faculdade de Engenharia da Universidade do Porto (FEUP). E-mail: jpa@fe.up.pt.}
\thanks{J. Barros: Mailing address: Same as J. Almeida, Visitors: Building I, Office I338. Affiliation: Instituto de Telecomunica\c{c}\~{o}es, Departamento de Engenharia Electrot\'{e}cnica e de Computadores, FEUP. Phone: (+351)225081825. E-mail: jbarros@fe.up.pt.}
\else
\fi}%


\ifCLASSOPTIONpeerreview
\markboth{Submission to IEEE Transactions on Information Forensics and Security}%
{Achieving Physical-Layer Security through Stopping Sets in a Feedback Wiretap Scenario} 
\else
\fi

\maketitle

\begin{abstract}

In this paper we discuss the ability of channel codes to enhance cryptographic secrecy. Toward that end, we present the secrecy metric of degrees of freedom in an attacker's knowledge of the cryptogram, which is similar to equivocation. Using this notion of secrecy, we show how a specific practical channel coding system can be used to hide information about the ciphertext, thus increasing the difficulty of cryptographic attacks. The system setup is the wiretap channel model where transmitted data traverse through independent packet erasure channels with public feedback for authenticated ARQ (Automatic Repeat reQuest). The code design relies on puncturing nonsystematic low-density parity-check codes with the intent of inflicting an eavesdropper with stopping sets in the decoder. Furthermore, the design amplifies errors when stopping sets occur such that a receiver must guess all the channel-erased bits correctly to avoid an expected error rate of one half in the ciphertext. We extend previous results on the coding scheme by giving design criteria that reduces the effectiveness of a maximum-likelihood attack to that of a message-passing attack. We further extend security analysis to models with multiple receivers and collaborative attackers. Cryptographic security is enhanced in all these cases by exploiting properties of the physical-layer. The enhancement is accurately presented as a function of the degrees of freedom in the eavesdropper's knowledge of the ciphertext, and is even shown to be present when eavesdroppers have better channel quality than legitimate receivers.

\end{abstract}


\ifCLASSOPTIONpeerreview
\begin{center} \bfseries EDICS Categories: SYS-PROT, SEC-MLEV, CRY-INTE, MOD-SECU, and CRY-CRYP \end{center}
\fi
\IEEEpeerreviewmaketitle

\section{Introduction}\label{sec:intro}

\subsection{Cryptography and the Physical Layer}

\IEEEPARstart{M}any cryptosystems in place today measure security computationally. If all known attacks are computationally intractable, then the system is deemed to be secure. The chief failings of this notion of security are the assumptions placed on the attacker. First, it is assumed that the attacker has limited resources to confront the problem, even if those resources are state of the art. Second, it is assumed that the attacker uses attacks which are publicly known, even though a better attack may exist. Claude Shannon addressed these shortcomings by defining the notion of \emph{perfect secrecy} \cite{Shannon49}. If a secret message $M$ is encrypted into a cryptogram $E$ using a secret key $K$, then perfect secrecy is achieved if $H(M|E)=H(M)$. Shannon also proved that perfect secrecy is only attainable if the key is at least as long as $M$, which is clearly impractical. However, perfect secrecy also makes the limiting assumption that an attacker has access to an error-free cryptogram, which may not be the case in practice.

Aaron Wyner later introduced the wiretap channel model, along with a new condition for secrecy \cite{Wyner75}. Let a message $M$ of length $k$ be encoded into a codeword $X$ of length $n$, and then transmitted. The rate of the encoder is $k/n$. A legitimate receiver obtains $Y$ over the \emph{main channel} denoted $Q_m$, and an eavesdropper obtains $Z$ over a \emph{wiretap channel} denoted $Q_w$. The secrecy condition is \begin{equation}\label{eq:weakSecrecy}
 \lim_{k\rightarrow\infty}\frac{I(M;Z)}{k}=0.
\end{equation}
Wyner showed that for rates up to the secrecy capacity $C_s$, encoders and decoders exist which can satisfy (\ref{eq:weakSecrecy}) and also achieve arbitrarily low probability of error for intended parties when $X\rightarrow Y\rightarrow Z$ is a Markov chain. This is known as the \emph{degraded} wiretap channel model. Csisz\'{a}r and K\"{o}rner \cite{Csiszar78} later generalized these results removing the degraded restriction, but still showing that $C_s > 0$, only if $Q_m$ is \emph{less noisy} than $Q_w$.

Understanding of the theoretically achievable secrecy rates of communication systems has continued to grow, as outlined in e.g. \cite{Wyner84}, \cite{Bloch06}, and \cite{Maurer00}. But another of the main challenges in this area has been the design of practical systems which achieve the secrecy rates indicated by the theory. These systems exploit noise in the channel at the physical layer of the communications system. Practical designs maximizing the information-theoretic secrecy are not trivial. Most currently suffer from one or more of several drawbacks. For instance, code designs are oftentimes a function of specific channel parameters (channel state information or CSI) seen by legitimate receivers and eavesdroppers.  Without accurate CSI, the results of these systems are not guaranteed; therefore, channels with varying or unknowable parameters present design issues. Other codes offer secrecy for only specific types of channels, or only when the eavesdropper's channel is degraded. Still other designs are impractical in the real world due to design complexity, necessary side information for legitimate decoding, or other limitations. Finally, the most glaring shortcoming of any scheme which derives security from the physical layer of a communications system, is that if an eavesdropper has a \emph{better} channel than a legitimate receiver, the scheme is likely to fail. The extreme case is when an eavesdropper has a noise-free channel and $Z=X$. Clearly this necessitates any physical-layer security scheme to be coupled with some other protection in order to maintain secrecy in the worst case.

\subsection{Main Contributions}

The intent of this paper is to develop the notion of \emph{combined security} due to cryptography and channel coding, thus providing a more complete security solution. To accomplish this goal, we cast coding into a cryptographic enhancement role, and seek to prevent an attacker from obtaining a noise-free cryptogram using channel coding. We present a new security metric for physical-layer schemes; namely, degrees of freedom $D$ in an attacker's knowledge of the cryptogram. As a comparison, if bits in $M$ are uniformly zero or one and independent and identically distributed (i.i.d.), then perfect secrecy implies $D=k$. In fact we show that $H(X|Z) = E[D]$ for a specific case. Our notion of physical-layer security using $D$ addresses the effectiveness of attacks on a cryptographic layer. To be more precise, our notion of security answers the practical question, how does the complexity of an attack on the cryptography change without perfect knowledge of the cryptogram?

It has been shown previously using correlation attacks on stream ciphers that certain cryptographic attacks are still possible even on noisy cryptograms, although a threshold on the noise level exists such that errors beyond the threshold cause the attack to fail \cite{Meier89, Harrison09_ICC, Harrison09_ISIT, Harrison09_ITW}. Practical schemes should provide enough confusion to exploit even the smallest amount of noise in an eavesdropper's received data to cause failure of these attacks on the cryptographic layer. Such systems should be robust to varying channel parameters, imperfect CSI at the encoder, and nondegraded system models. In fact, good designs still offer security enhancement to cryptography, even when attackers have an advantage in signal quality over legitimate receivers. Of course, all of this must be done while guaranteeing reliable communication between friendly parties.

Therefore, along with the new metric, this paper also analyzes combined cryptographic and physical-layer security in a \emph{practical coding scheme} using degrees of freedom to characterize security. In \cite{Harrison10_ITW}, this scheme was shown to inflict a passive eavesdropper using a message-passing decoder with stopping sets with very high probability when a legitimate receiver and an eavesdropper view transmitted data through statistically independent packet erasure channels (PEC). The scheme relies on a nonsystematic low-density parity-check (LDPC) code design, with puncturing and interleaving steps in the encoder. Legitimate receivers are given access to an authenticated public feedback channel for Automatic Repeat-reQuest (ARQ). In this paper, we broaden the security analysis of the scheme given in \cite{Harrison10_ITW} by addressing the following points.
\begin{itemize}
  \item \emph{Degrees of Freedom:} The system security is analyzed using the new metric. Computational secrecy is shown to grow exponentially with $E[D]$, which is also shown to be equal to $H(X|Z)$ for the prescribed encoder.
  \item \emph{Encoder Description:} End-to-end details of the encoder and decoder are provided, as well as simulation results which match theoretical expectations.
  \item \emph{Optimization:} Design criteria are specified to maximize the degrees of freedom in the maximum-likelihood attack as well as the message-passing attack. This involves comparison of irregular LDPC codes with regular LDPC codes.
  \item \emph{Extensions:} Security results are made general so as to apply to multiple receivers and multiple collaborative attackers. Ultimately, bounds on the increase in computational secrecy of an underlying cryptosystem are specified when the physical-layer encoding system is employed.
\end{itemize}
Ultimately, this scheme has very few design constraints, offers enhanced cryptographic secrecy over a wide range of CSI parameters, and requires no secret key and no rate reduction in data transmission.

\subsection{Related Works}

Our encoder makes use of fundamental practical design ideas which have been shown to offer secrecy. For example, our encoder employs nonsystematic LDPC codes in order to hide information bits and magnify coding errors. Secrecy properties of these codes have been studied in \cite{Baldi10}. We further employ intentional puncturing of encoded bits, a technique shown to offer security in \cite{KlincITW09, Klinc09GC}. Our scheme punctures with the goal of inducing \emph{stopping sets} in an eavesdropper's received data. As a result, every transmitted bit is crucial for decoding. Our intent is to punish an eavesdropper for every missing piece of information. Finally, in order to distribute erasures throughout the data set, the encoder interleaves coded bits among several transmitted packets. Similar ideas of interleaving coded symbols have been used in \cite{Bloch08, MANETISIT09} in conjunction with wiretap codes developed in \cite{Thangaraj07} to offer secrecy to various systems. The works \cite{Lai08, Latif09} give results for ARQ and feedback wiretap systems.

It can be argued that the first practical secrecy coding scheme was presented by Ozarow and Wyner in an extension of the original wiretap paper \cite{Wyner84}. Here the general idea of partitioning a group code into cosets to achieve secrecy was first presented. This technique was shown to apply to LDPC codes much more recently in \cite{Thangaraj07}, and achieves the secrecy condition in (\ref{eq:weakSecrecy}) for noiseless main channels when the wiretap channel is either a binary erasure channel (BEC) or a binary symmetric channel (BSC). This work in LDPC codes for secrecy has been furthered in \cite{Arun10_ITW}, where large-girth LDPC codes are considered, and shown to meet the secrecy constraint in (\ref{eq:weakSecrecy}) for noiseless main channel and BEC wiretap channel. A stronger notion of secrecy than (\ref{eq:weakSecrecy}) is also achieved for these codes in certain cases. Finally, it should be noted that Ar{\i}kan's polar codes \cite{Arikan09} can offer secrecy for general symmetric channels, although code construction is an issue for non-erasure channels. Schemes have been presented in \cite{Hof10} and \cite{Mahdavifar10} which achieve the secrecy capacity under the condition in (\ref{eq:weakSecrecy}), although these schemes only offer secrecy for degraded wiretap channels. Furthermore, design of these codes is heavily contingent on perfect CSI at the encoder.

Although our codes can be shown to achieve (\ref{eq:weakSecrecy}) only under certain puncturing criteria, the main contribution of the coding scheme presented here is the cryptographic security enhancements shown using degrees of freedom as a security metric. Our scheme is robust against imperfect CSI, and for that matter, undetected eavesdroppers. According to our knowledge, it is also the first \emph{practical} secrecy scheme which can operate on the general wiretap channel (nondegraded case) when both $Q_m$ and $Q_w$ are erasure channels.

The rest of the paper is outlined as follows. In Section \ref{sec:system}, we discuss the system model for which our encoder is designed, which is an adaptation of the wiretap channel model from \cite{Wyner75}. The precise definition of degrees of freedom is also given. Section \ref{sec:stoppingSets} addresses background information regarding LDPC codes and stopping sets. Our novel encoder and decoder designs are presented in Sections \ref{sec:encoder} and \ref{sec:decoder}, respectively. Analysis of the security inherent in the system is then completed in Section \ref{sec:analysis} for various scenarios, ultimately culminating in the most general case which encompasses multiple users and collaborating eavesdroppers. Finally, bounds regarding enhancements of cryptographic security are presented in Section \ref{sec:crypto} along with end-to-end simulations of the system. Conclusions are provided in Section \ref{sec:conclusion}.

\section{System Model and Degrees of Freedom}\label{sec:system}

\begin{figure}
\begin{center}
  \begin{tikzpicture}
    [node distance=0.8cm, rounded corners=2pt, channel/.style={rectangle,draw=blue,fill=blue!10,thick,
    text centered,minimum size=8mm},
    boxedNode/.style={rectangle,draw,fill=black!10,thick,text centered, minimum size=8mm},
    inner sep=1mm]
    \node (Alice) {Alice};
    \node [boxedNode] (Encoder) [right=of Alice] {Encoder};
    \node [channel] (Qm) [right=of Encoder,label=above:$Q_m$] {PEC($\delta$)};
    \node [boxedNode] (Decoder) [right=of Qm] {Decoder};
    \node (Bob) [right=of Decoder] {Bob};
    \node [channel] (Qw) [below=of Qm,label=above:$Q_w$] {PEC($\epsilon$)};
    \node [boxedNode] (Decoder_w) [right=of Qw] {Decoder};
    \node (Eve) [right=of Decoder_w] {Eve};
    \draw[->] (Alice) -- node [above] {$M$} (Encoder);
    \draw[->] (Encoder) -- node [above] {$X$} (Qm);
    \draw[->] (Qm) -- node [above] {$Y$} (Decoder);
    \draw[->] (Decoder) -- node [above] {$\tilde{M}$} (Bob);
    \draw[->] (Qw) to node [above] {$Z$} (Decoder_w);
    \draw[->] (Decoder_w) to node [above] {$\hat{M}$} (Eve);
    \draw[->] ($(Encoder.east) + (4mm,0)$) |- (node cs:name=Qw,anchor=west);
    \draw [->] (Bob) -- ++(0,1.1)
        -| node[near start, above] {Feedback Channel} (Encoder);
  \end{tikzpicture}
\end{center}
\caption{Wiretap channel model with feedback assuming packet erasure channels for both the main channel $Q_m$ and the wiretap channel $Q_w$.}\label{fig:wiretapChannel}
\end{figure}
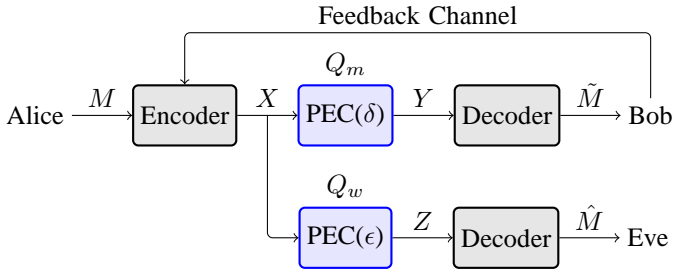

We begin by presenting the wiretap channel model \cite{BlochBook} with the addition of feedback in Fig. \ref{fig:wiretapChannel}. A user named Alice wishes to transmit an \emph{encrypted} binary message $M = (m^1, m^2, \ldots, m^L)$ to a legitimate receiver named Bob, where $m^i = (m_1^i, m_2^i, \ldots, m_k^i) \in \mathcal{M}$ for $i = 1,2,\ldots,L$. It will be helpful to think of $M$ as being broken up into $L$ blocks of length $k$, where $k$ is the dimension of the encoder to follow. The final block $m^L$ can be filled by concatenating random bits if needed. Let us also define the blocklength $n$ of the encoder. Then the coding rate is $k/n$. To be clear, $n$ is the length of a codeword after it has been punctured. We will also assume that $M$ has been compressed, so that all possible bit combinations are equally likely in the alphabet $\mathcal{M}$. Prior to transmission, Alice encodes $M$, resulting in a collection of $\eta$ packets $X = (x^1, x^2, \ldots, x^\eta)$ for transmission. Bob receives the packets as $Y$ through $Q_m$, a PEC with probability of erasure $\delta$. An eavesdropper named Eve obtains the packets $Z$, although through $Q_w$, an independent PEC with probability of erasure $\epsilon$. An obvious extension of this model is to consider correlated erasures in $Q_m$ and $Q_w$; however, in this paper we always assume erasures are statistically independent. Finally, $\tilde{M}$ and $\hat{M}$ are the respective estimates of $M$ by Bob and Eve.

The encoder and decoder exploit the independent nature of erased packets across $Q_m$ and $Q_w$. Of course, the system must guarantee that $\tilde{M} = M$, while at the same time making Eve as ignorant as possible. The authenticated feedback channel available to Bob plays a key role in accomplishing both of these endeavors. This public noiseless channel is used to request the retransmission of erased packets. Since it is authenticated, Alice is able to deduce whether Bob sent the request, and can detect any tampering with the data \cite{Stinson_CryptoBook}, which restricts Eve to passive status \cite{MaurerPart1}. Requests by Bob are public, and there is no \emph{secret key} employed at the physical layer. The sole source of confusion for Eve is her own naturally occurring erasure pattern across $Q_w$.\footnote{It is noted that results in Section \ref{sec:analysis} are provided for this system, as well as the more general model which allows an arbitrary number of legitimate receivers and eavesdroppers.}

As mentioned in Section \ref{sec:intro}, we define \emph{physical-layer security} for this system with the cryptographic layer in mind. Cryptographic attacks often assume an attacker has the luxury of an error-free version of $M$ (or even some of the plaintext), but our design aims to prevent this from occurring, by creating degrees of freedom in the attacker's knowledge of $M$.
\begin{definition}\label{def:degsFreedom}
 The number of \emph{degrees of freedom} in a received codeword is a random variable $D$ which takes on the number of encoded symbols for which an eavesdropper has no information. Therefore, the probabilities of all symbol values on these $D$ symbols are equally likely.
\end{definition}
For binary codes with $D=d$, a codeword of length $n$ can be any of $2^d$ equally likely codewords, each mapping to a unique $k$-bit message in $\mathcal{M}$. Since we assume that the attacker knows the encoder, the maximum value of $D$ is $k$, and can be shown to have an information-theoretic definition. Since an attacker has no knowledge of these bits, an average of $2^{E[D]-1}$ guesses must be made to obtain them. Using this reasoning, the goals of our physical-layer design are: first, to ensure that $D=0$ for Bob so that $\tilde{M} = M$; second, to make $D$ as large as possible for Eve; and third, to ensure that attacks on the cryptogram fail if $\hat{M} \neq M$.

\section{LDPC Codes and Stopping Sets}\label{sec:stoppingSets}
We employ LDPC codes \cite{Gallager63} and exploit the phenomenon of stopping sets to obtain security from the physical layer. This section provides limited background of LDPC codes and stopping sets in order to establish the foundation upon which to present our encoder.

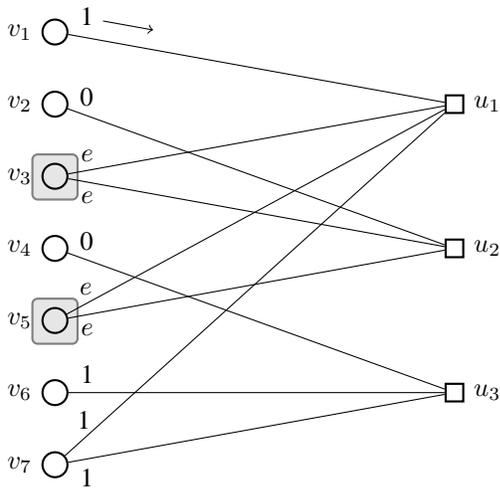
\begin{figure}
\begin{center}
  \begin{tikzpicture}
   [variableNode/.style={circle,draw,thick},
   stopNode/.style={circle,draw,thick},
   stopBlock/.style={rectangle,rounded corners=2pt,fill=black!10,thick,draw=black!50,inner sep=3mm,yshift=1.25mm},
   checkNode/.style={rectangle,draw,thick}, node distance=6mm]
    \node [variableNode] (v_one) [label=left:$v_1$] {};
    \node [variableNode] (v_two) [below=of v_one,label=left:$v_2$] {};
    \node [stopBlock] [below=of v_two] {};
    \node [stopNode] (v_three) [below=of v_two,label=left:$v_3$] {};
    \node [variableNode] (v_four) [below=of v_three,label=left:$v_4$] {};
    \node [stopBlock] [below=of v_four] {};
    \node [stopNode] (v_five) [below=of v_four,label=left:$v_5$] {};
    \node [variableNode] (v_six) [below=of v_five,label=left:$v_6$] {};
    \node [variableNode] (v_seven) [below=of v_six,label=left:$v_7$] {};
    \begin{scope}[node distance=5cm]
    \node [checkNode] (u_one) [right=of v_two,label=right:$u_1$] {};
    \node [checkNode] (u_two) [right=of v_four,label=right:$u_2$] {};
    \node [checkNode] (u_three) [right=of v_six,label=right:$u_3$] {};
    \end{scope}

    \draw (v_one) -- node[pos=0.05, above,text height=1.5ex, text depth=.25ex] (messageNode1) {1} node[pos=0.25, above,text height=1.5ex,text depth=.25ex] (arrowNode1) {} (u_one);
    \draw [->,xshift=5pt,yshift=5pt] (messageNode1) -- (arrowNode1);
    \draw (v_two) -- node[pos=0.05, above] {0} (u_two);
    \draw (u_one) -- node[pos=0.95, above] {$e$} (v_three) -- node[pos=0.05, below] {$e$} (u_two);
    \draw (v_four) -- node[pos=0.05, above] {0} (u_three);
    \draw (u_one) -- node[pos=0.95, above] {$e$} (v_five) -- node[pos=0.05, below] {$e$} (u_two);
    \draw (v_six) -- node[pos=0.05, above] {1} (u_three);
    \draw (u_one) -- node[pos=0.95, above] {1} (v_seven) -- node[pos=0.05, below] {1} (u_three);

  \end{tikzpicture}
\end{center}
\caption{Tanner graph for MP decoding over the BEC with a highlighted stopping set due to erasures at variable nodes $v_3$ and $v_5$.}\label{fig:stoppingSet}
\end{figure}

Let us define a general binary LDPC code $C$ with blocklength $N$, and dimension $k$. Note that this $k$ is identical to $k$ from section \ref{sec:system}, but $N$ the blocklength of the LDPC code, is different from $n$ the blocklength of the encoder because $n$ is the codeword length after puncturing. The parity check matrix $H$ fully defines the code, and is $N - k \times N$. We will find it helpful to think of $H$ in terms of its corresponding Tanner graph $G_C$ \cite{MoonArches,UrbankeModern}. The set of variable nodes is $V = (v_1, v_2, \ldots, v_N)$, while the set of check nodes is $U = (u_1, u_2, \ldots, u_{N-k})$. Variable nodes correspond to the $N$ bits in a codeword. Checks correspond to rows in $H$, where the set of bits that participate in the check $u_i$ is denoted $\mathcal{N}_i = \{j: H_{i,j} = 1\}$ \cite{MoonArches}. Then the $i$th check is calculated in GF(2) as $u_i = \sum_{j\in \mathcal{N}_i}v_j = 0$. The notation $\mathcal{N}_{i,j}$ signifies all bits in the $i$th check except the $j$th bit. The $j$th variable node shares an edge with the $i$th check node in $G_C$ if and only if $j\in \mathcal{N}_i$. The Tanner graph for a simple example is shown in Fig. \ref{fig:stoppingSet}.

Decoding of an LDPC codeword over a BEC can be accomplished using maximum-likelihood (ML) decoding \cite{Burshtein04}, by solving a system of equations. However, the iterative message-passing (MP) decoder is commonly used due to its computational efficiency. We briefly explain both decoders.

\subsection{Maximum-Likelihood Decoding}\label{subsec:ML}
Let us consider an LDPC codeword $x\in C$ transmitted over a BEC and let $y$ denote the received codeword. Note that $x_i\in\{0,1\}$ and $y_i\in\{0,1,e\}$ where $e$ signifies an erased bit. We let $\mathcal{K}$ denote the set of known bits in $y$, and $\bar{\mathcal{K}}$ denote the set of erased bits in $y$. Furthermore, $H_{\mathcal{K}}$ and $H_{\bar{\mathcal{K}}}$ can be understood to be matrices formed by the columns of $H$ indexed by $\mathcal{K}$ and $\bar{\mathcal{K}}$, respectively. Similarly, $x_{\mathcal{K}}$ and $x_{\bar{\mathcal{K}}}$ are vectors composed of only the bits indexed by the respective sets $\mathcal{K}$ and $\bar{\mathcal{K}}$.

Clearly, $0 = Hx^T = H_{\mathcal{K}}x_{\mathcal{K}}^T + H_{\bar{\mathcal{K}}}x_{\bar{\mathcal{K}}}^T$, where $x_{\mathcal{K}} = y_{\mathcal{K}}$, and thus $H_{\mathcal{K}}x_{\mathcal{K}}^T = z^T$ is known. The maximum likelihood decoder must then solve for the channel-erased bits $x_{\bar{\mathcal{K}}}$ using the system of equations given by
\begin{equation}\label{eq:systemML}
  H_{\bar{\mathcal{K}}}x_{\bar{\mathcal{K}}}^T = z^T.
\end{equation}
This system has a unique solution when the erased bits are such that the columns of $H_{\bar{\mathcal{K}}}$ are linearly independent \cite{Urbanke01}. We can obtain a bound from this statement which we will use to analyze security in the worst-case.
\begin{proposition}\label{prop:boundOnSizeKbar}
For a linear code $C$ with blocklength $N$ and dimension $k$, the ML decoder over the BEC cannot have a unique solution if the number of erasures exceeds $N - k$, that is if $|\bar{\mathcal{K}}| > N - k$.
\end{proposition}
\begin{IEEEproof}
The rank of $H_{\bar{\mathcal{K}}}$ equals the number of linearly independent rows or columns of the matrix (\cite{MoonBlack}, pg. 244). Since $N - k$ is the number of rows in $H$, the rank of $H_{\bar{\mathcal{K}}}$ can never exceed $N - k$, and thus the ML decoder cannot produce a unique solution when $|\bar{\mathcal{K}}| > N - k$.
\end{IEEEproof}
In fact, when the number of erasures exceeds $N - k$, the system in (\ref{eq:systemML}) will be such that the degrees of freedom in the ML decoder $D_{ML} \geq |\bar{\mathcal{K}}| - (N - k)$, where we achieve equality if there are $N - k$ linearly independent columns in $H_{\bar{\mathcal{K}}}$ \cite{Burshtein04}. In any case, $D_{ML}$ is equal to the difference in the number of erased bits, and the number of linearly independent columns of $H_{\bar{\mathcal{K}}}$, and is zero if this difference is negative. This definition clearly satisfies the notion of degrees of freedom from Definition \ref{def:degsFreedom} for this decoder. Thus we see that the effectiveness of the decoder is strictly bounded by the redundancy of the code. While faster methods have been discovered for solving a linear system of equations, the straightforward decoder is known to have complexity $((1-R)\beta+\gamma \delta)\delta^2N^3$, where $R$ is the rate of the code, $\beta$ and $\gamma$ are constants which are also a function of the elimination algorithm chosen to solve the system of equations, $\delta$ is the erasure probability in the channel, and $N$ is the blocklength of the code \cite{Burshtein04}.

\subsection{Message-Passing Decoding}\label{subsec:MP}

Let $C$, $x$, and $y$ hold the same definitions as for the ML decoder. The MP decoder is an iterative decoder based on the Tanner graph representation of $C$. The decoding process passes \emph{messages} between $U$ and $V$ along the edges of $G_C$. One version of the decoder is given as Algorithm \ref{alg:MPdecoding} (adapted from \cite{Urbanke01}). The number of degrees of freedom in the MP decoder $D_{MP}$ is the cardinality of the smallest set of bit values that must be supplied in order to decode all remaining bits. If the decoder succeeds, then $D_{MP} = 0$. Clearly, this maintains the definition of degrees of freedom given in Definition \ref{def:degsFreedom} when restricted to this decoder, because any bit combination of these $D_{MP}$ values decodes to a valid codeword, and each is equally likely without further information. A bound on the correction capabilities of the MP decoder is given by the following proposition.
\begin{proposition}\label{prop:maxDecodMP}
  The MP decoder over the BEC can correct no more than $N-k$ erasures.
\end{proposition}
\begin{IEEEproof}
 In Algorithm \ref{alg:MPdecoding}, each check node can correct at most one variable node, and $|U|=N-k$.
\end{IEEEproof}
The MP decoder is suboptimal compared with the ML decoder, although the MP decoder has linear complexity in the blocklength \cite{MoonArches}. A more detailed comparison of the two decoders is offered in \cite{Lee07}.

\algsetup{indent=2em}
\begin{algorithm}[h]
\caption{Message-Passing Decoder over the BEC \cite{Urbanke01}.}
\begin{algorithmic}[1]
\STATE \textbf{Initialize:} For $y_i \neq e$, set $v_i = y_i$ and declare all such variable nodes as known.
  \IF {(No variable nodes are known and no check node has degree one)}
    \STATE Output the (possibly partial) codeword and stop.
  \ELSE
    \STATE Delete all known variable nodes along with their adjacent edges.
  \ENDIF
  \STATE For each variable node $v_j$ connected to a degree one check node $u_i$, declare $v_j$ as known and set $v_j = \sum_{k\in \mathcal{N}_{i,j}}v_k$. Jump to 2.
\end{algorithmic}
\label{alg:MPdecoding}
\end{algorithm}

\subsection{Stopping Sets}\label{subsec:stoppingSets}

In order to make $D$ as large as possible for our system when an eavesdropper uses an MP decoder, we would like to design the encoder block from Fig. \ref{fig:wiretapChannel} so that every bit erased by the channel adds a degree of freedom to the decoder. Stopping sets provide a means of accomplishing this task.
\begin{definition}[Di, et. al. \cite{Di02StoppingSets}]\label{def:stoppingSet}
 A \emph{stopping set} is a set $S \subseteq V$ such that all check nodes in $N(S)$ are connected to $S$ by at least two edges, where $N(S)$ signifies the \emph{neighborhood} of $S$ and is defined as the set of all adjacent nodes to any member of $S$ in $G_C$.
\end{definition}
Notice that the empty set, by definition, is a stopping set, as is any union of stopping sets. Thus, any set of variable nodes has a unique maximal stopping set in it.\footnote{For our purposes, we will sometimes ignore the empty set as a stopping set and say that a set $A$ \emph{contains no stopping sets}, meaning that the maximal stopping set in $A$ is $\emptyset$.} See Fig. \ref{fig:stoppingSet} for a simple example; clearly the erasures cannot be resolved using Algorithm \ref{alg:MPdecoding}. This gives way to the following lemma, proved in \cite{Di02StoppingSets}.
\begin{lemma}[Di et. al. \cite{Di02StoppingSets}, Lemma 1.1] \label{lem:stoppingSets}
   Let $G$ be the Tanner graph defined by the parity check matrix $H$ of a binary linear block code $C$, and assume that $C$ is used to transmit over the BEC. Let $A$ be the set of erased bits in the received codeword. Then, using Algorithm \ref{alg:MPdecoding} on $G$, the set of erasures which remain after decoding comprise the unique maximal stopping set in $A$.
\end{lemma}

Since stopping sets cause the MP decoder to fail, puncturing in the encoder will be done with an attempt to inflict Eve with stopping sets. However, the ML decoder will still succeed, even in the presence of stopping sets, as long as the erased bits have linearly independent columns in $H$. We account for both decoders in our design by using a particular ensemble of LDPC codes where $D_{MP}$ can be made equal to $D_{ML}$, thus ensuring secrecy regardless of the decoder used by Eve. The simplicity of MP decoding is also preserved for all legitimate receivers.\footnote{For further information on stopping sets as they relate to LDPC code ensembles, see \cite{Rosnes09} and \cite{Orlitsky05}.}

\section{Encoder}\label{sec:encoder}

\begin{figure}
\begin{center}
  \begin{tikzpicture}
  [node distance=0.35cm, rounded corners=2pt, boxedNode/.style={rectangle,draw,fill=black!10,thick,
  text centered, minimum size=8mm},
  multiLine/.style={rectangle,text centered, text width=1.3cm, minimum size=8mm},
  boxMultiLine/.style={rectangle,draw,fill=black!10,thick,
  text centered, text width=1.3cm, minimum size=8mm},
  inner sep=1mm]
    \node [boxMultiLine] (Encoder) {LDPC Encoder};
    \node [boxMultiLine] (Puncture)  [right=of Encoder]  {Puncture Block};
    \node [boxedNode] (Buffer)  [right=of Puncture]  {Buffer};
    \node [boxedNode] (Interleaver)  [right=of Buffer]  {Interleaver};
    \draw[->] ($(Encoder.west) + (-3mm,0)$) -- node [midway, above] {$M$} node [midway, below=14pt, text width=1.5cm, text centered] {$L$ blocks length $k$} (Encoder);
    \draw[->] (Encoder) to node [above, midway] {$B$} node [below=14pt, text width=1.5cm, text centered, midway] {$L$ blocks length $N$} (Puncture);
    \draw[->] (Puncture) to node [above] {$P$} node [below=14pt, text width=1.5cm, text centered, midway] {$L$ blocks length $n$} (Buffer);
    \draw[->] (Buffer) to (Interleaver);
    \draw[->] (Interleaver) to node [midway, above] {$X$} node [midway, below=12pt, text width=1.6cm, text centered] {$\eta$ packets size $\alpha L$} ($(Interleaver.east) + (3mm,0)$);
  \end{tikzpicture}
\end{center}
\caption{Detailed block diagram of the encoder. Number and size of blocks or packets are indicated at each step.}\label{fig:encoder}
\end{figure}
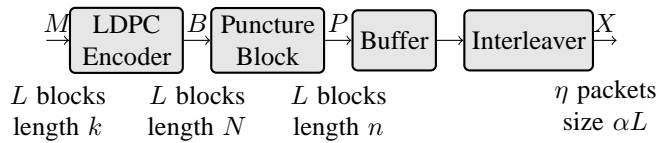

The encoder design is based on the fact that $I(M;Z) \leq I(M;X)$ because processing cannot increase information, and $M\rightarrow X\rightarrow Z$ is a Markov process \cite{Cover}. The key idea in the decoder is to reduce $X$ to the decoding threshold. In other words, $X$ can be used to recover $M$ by design, but if any erasures remain in $Z$ following transmission, unique decodability is not possible. Proper design maximizes $D$ for Eve. The stages of encoding are portrayed in Fig. \ref{fig:encoder}, where each stage fulfills a specific purpose within the overall goals of obtaining secrecy and reliability. The following principles are addressed in the design of this encoder.
\begin{itemize}
  \item Bits of $M$ are hidden from immediate access in the decoded words using nonsystematic LDPC codes.
  \item Scrambling prior to coding magnifies errors due to the physical layer of the communication system.
  \item The error-correction capabilities of the LDPC code are restricted by intentional puncturing of encoded bits. (Bob obtains reliability through ARQ, rather than error correction.)
  \item Bits from encoded blocks are interleaved amongst several transmitted packets so that a single erased packet results in erasures in many encoded blocks of data.
\end{itemize}

\begin{singlespace}
\end{singlespace}

\subsection{Nonsystematic LDPC Codes}\label{subsec:nonsystematic}

Recall from Section \ref{sec:system} that $M = (m^1, m^2, \ldots, m^L)$, where $m^i = (m_1^i, m_2^i, \ldots, m_k^i)\in\mathcal{M}$ for $i = 1,2,\ldots,L$. These $L$ blocks of encrypted message form the input to the nonsystematic LDPC encoder with blocklength $N$ and dimension $k$. The output of the LDPC encoder $B$ is given as $L$ codewords of length $N$, denoted as $B = (b^1, b^2, \ldots, b^L)$ where each vector $b^i = (b_1^i, b_2^i, \ldots, b_N^i)$. Certainly, if the code $C$ were systematic, then the bits of $m^i$ would appear explicitly in the encoded block $b^i$. For secrecy purposes, nonsystematic codes are employed.

Nonsystematic LDPC coding is typically implemented as a two stage process to improve encoder complexity \cite{Alloum05, Shamir05, Baldi10}. Let $S$ be an invertible $k \times k$ \emph{scrambling} matrix in GF(2), and let $G$ be a $k \times N$ systematic generator matrix. Let $m$ be a length-$k$ message. Then our LDPC encoding process applies the scrambling matrix to $m$ as
\begin{equation}\label{eq:scramble}
  m' = mS.
\end{equation}
The data are then encoded using $G$ by $b = m'G$ to obtain a length-$N$ block of encoded data. Clearly at the decoder the inverse operation first requires the bits of $b$ to be obtained through either MP or ML decoding. Since $G$ is systematic, the bits of $m'$ are explicit in $b$. The bits of $m$ can then be found by applying the inverse of $S$ in the descrambling operation
\begin{equation}\label{eq:inverseScramble}
  m = m'S^{-1}.
\end{equation}
This process amplifies errors in the decoding process as a function of the sparsity of $S^{-1}$. Note that $S^{-1}$ can be obtained through e.g. LU decomposition \cite{MoonBlack}, with modifications for GF(2). In our experience, randomly generated scrambling matrices which are nonsingular are likely to have inverses with just less than 50\% of the entries equal to one on average. If $S$ matrices are randomly generated until one can be inverted to obtain $S^{-1}$, the resulting despreading operation is enough to cause even a single error in $m'$ to result in roughly a 50\% error rate in $m$ as shown in Section \ref{sec:crypto}. Although this can be made more precise, the result is intuitive because a bit in $m$ is a linear combination of bits in $m'$. Thus, if there are an odd number of bits in error in a given combination of say $m_i$, then that bit will be in error. On average, the row weight in $S^{-1}$ is approximately $k/2$, and the expectation of $k/2$ bits in error holds for any number of errors in $m'$.

Since only one $(S,S^{-1})$ pair need be used by the system, the matrices can be generated off-line, which does not affect encoding and decoding complexity. However, the complexity of both the encoder and the decoder is increased due to the matrix multiplications in (\ref{eq:scramble}) and (\ref{eq:inverseScramble}). Both of these operations are $\mathcal{O}(k^3)$. General systematic encoder complexity is $\mathcal{O}(N^2)$ because $G$ is not sparse by design \cite{MoonArches}, although improvements can be made using appropriate preprocessing as outlined in \cite{Urbanke01}. The encoding technique specified in \cite{Urbanke01} gives encoder complexity of $\mathcal{O}(N + g^2)$ where $g$ is the \emph{gap} in an approximate lower triangular form of the parity check matrix and is less than $N - k$. The complexities for the ML and MP decoders are given in Sections \ref{subsec:ML} and \ref{subsec:MP} as $\mathcal{O}(N^3)$ and $\mathcal{O}(N)$, respectively.

\subsection{Puncturing}\label{subsec:puncturing}

The next step in the encoding process is to apply a puncturing pattern to each codeword in $B$. Let the puncturing pattern $R\in V$ indicate which bits in each $b^i$ are to be punctured. Recall that $V$ is the set of variable nodes in the Tanner graph $G_C$. The punctured blocks $P = (p^1, p^2, \ldots, p^L)$, where each $p^i = (p_1^i, p_2^i, \ldots, p_{n}^i)$ are shown in Fig. \ref{fig:encoder} to have length $n$, which was defined in Section \ref{sec:system} to be the blocklength of the encoder. All bits which are not punctured belong to the set $Q$ so that $V = R + Q$; therefore, the length of each block in $P$ is equal to $|Q|=n$. The puncturing pattern is chosen in order to induce stopping sets in an eavesdropper's received data.
\begin{definition}\label{def:acceptable}
 A puncturing pattern $R$ is deemed \emph{acceptable} if and only if there are no stopping sets in $R$, and $R + v$ contains some nonempty stopping set $S_v$ for every variable node $v \in Q$.
\end{definition}
Such a set $R$ can be constructed using the random technique outlined in Algorithm \ref{alg:findR}, which also calls Algorithm \ref{alg:isThereStoppingSet} in order to check for stopping sets in a computationally tractable manner \cite{Harrison10_ITW}.

\algsetup{indent=2em}
\begin{algorithm}[h]
\caption{Finds an acceptable puncturing pattern $R$ within the set of all variable nodes $V$.}
\begin{algorithmic}[1]
\STATE \textbf{Initialize:} $R = v$, for a randomly chosen $v\in V$, and $Q = \emptyset$.
  \IF{ ($V\backslash(R\cup Q) \neq \emptyset$)}
    \STATE Choose another $v$ randomly from $V\backslash(R\cup Q)$.
    \STATE Run Algorithm \ref{alg:isThereStoppingSet} with $A = R + v$ to check for stopping sets.
    \IF{($R+v$ has a stopping set, i.e. Algorithm \ref{alg:isThereStoppingSet} returns true)}
      \STATE $Q = Q + v$.
    \ELSE
        \STATE $R = R + v$.
    \ENDIF
    \STATE Jump to 2.
  \ELSE
    \STATE Terminate.
  \ENDIF
\end{algorithmic}
\label{alg:findR}
\end{algorithm}

\algsetup{indent=2em}
\begin{algorithm}[h]
\caption{Checks for the existence of stopping sets in a subset of variable nodes, $A\subseteq V$ \cite{Harrison10_ITW}.}
\begin{algorithmic}[1]
\STATE \textbf{Initialize:} $S = A$
  \IF { $(S \neq \emptyset)$ }
    \STATE Induce subgraph $G'$ in $G$ using ($S \cup N(S))$.
    \IF{ ($\exists$ a check node in $G'$ with degree 1)}
      \STATE Delete variable nodes from $S$ which are adjacent to check nodes of degree 1 in $G'$, jump to 2.
    \ELSE
      \STATE Return true. $S$ is the maximal nonempty stopping set in $A$.
    \ENDIF
  \ELSE
    \STATE Return false. There is no nonempty stopping set in $A$.
  \ENDIF
\end{algorithmic}
\label{alg:isThereStoppingSet}
\end{algorithm}

\begin{lemma}\label{lem:algorithmFindR}
 The output of Algorithm \ref{alg:findR} is always an \emph{acceptable} puncturing pattern R as defined in Definition \ref{def:acceptable}.
\end{lemma}
\begin{IEEEproof}
We must first show that upon completion of Algorithm \ref{alg:findR}, there are no stopping sets in $R$. Assume for a contradiction that $R$ has a stopping set. Then there is a bit $v\in R$ which when added to $R$ during the construction process, caused a stopping set to first appear. Then by Algorithm \ref{alg:findR}, $v\notin R$. This provides the contradiction. It remains to be proved that Algorithm \ref{alg:isThereStoppingSet} operates as expected.
\begin{proposition} \label{prop:algIsThereSSWorks}
 Algorithm \ref{alg:isThereStoppingSet} always returns true when $A$ has a nonempty stopping set, and always returns false otherwise.
\end{proposition}
\begin{IEEEproof}[Proof of Proposition] Suppose that the bits in $A$ were actually erasures over the BEC, and Algorithm \ref{alg:MPdecoding} was used to decode. Realize that erasures recovered in the $i$th iteration of Algorithm \ref{alg:MPdecoding} correspond exactly to the nodes deleted in the $i$th iteration of Algorithm \ref{alg:isThereStoppingSet}. If all bits can be resolved using MP decoding then all nodes will be deleted in Algorithm \ref{alg:isThereStoppingSet}, and false is returned. If, however, MP decoding returns a partial codeword, then Algorithm \ref{alg:isThereStoppingSet} will return true because all remaining bits have degree greater than one in the induced subgraph $G'$. Therefore, by Lemma \ref{lem:stoppingSets}, the remaining nodes comprise the maximal stopping set of $A$.
\end{IEEEproof}

To complete the proof of Lemma \ref{lem:algorithmFindR}, we must also show that for any $v\in Q$, $R+v$ has a nonempty stopping set. Since in Algorithm \ref{alg:findR} every $v\in Q$ is such that for some subset $R'\subseteq R$, $R'+v$ has a stopping set, therefore $R + v$ has a stopping set for any $v\in Q$.
\end{IEEEproof}
Thus, puncturing according to $R$ in each $b^i$ for $i = 1, 2, \ldots, L$, guarantees that every bit in each $p^i$ is crucial for successful MP decoding.

Complexity of Algorithm \ref{alg:findR} is linear in the blocklength $N$, because it chooses $N-1$ bits in a random order, and calls Algorithm \ref{alg:isThereStoppingSet} after each choice. The complexity of Algorithm \ref{alg:isThereStoppingSet} in the worst case, is quadratic in $|U|=N-k$ the number of check nodes in $G_C$. Line 5 of the algorithm will be repeated a maximum of $\sum_{i=1}^{|U|} i = \frac{|U|^2+|U|}{2}$ times if a single node is deleted each time the line is executed. Therefore, the complexity of finding an acceptable puncturing pattern $R$ is at most quadratic in $|U|$, and linear in $N$, i.e. has complexity $\mathcal{O}(N|U|^2)$. Thus the algorithm can be used in practical system design to compute $R$ off-line.

\subsection{Regular vs. Irregular Codes}\label{subsec:regVsIrreg}

The overall rate $k/n$ of the nonsystematic and punctured code is a function of the rate of the systematic LDPC code, and $|R|$. Simulations have shown that the size of $R$ is very much a function of the degree distribution on $C$, although the exact relationship is still unknown.
\begin{example}\label{ex:punct}
Let $C$ be a regular rate-1/2 code with $N = 1000$, $w_c = 4$, and $w_r = 8$, where $w_c$ and $w_r$ are the fixed column and row weights of the parity check matrix, respectively. The size of $|R|$ appears to be Gaussian-distributed for this family of codes with a mean size of approximately 436, with variance roughly equal to 15. Let us examine, however, an irregular ensemble with the same rate and blocklength, but having the following edge degree distribution pair: $\eta(x) = 0.32660x + 0.11960x^2 + 0.18393x^3 + 0.36988x^4$ on variable node weights, and $\chi(x) = 0.78555x^5 + 0.21445x^6$ on check node weights (see \cite{MoonArches} pg. 664), where $H$ is formed using the socket approach given in \cite{Burshtein04}. Here the distribution on $|R|$ is much tighter, ranging from 496 to 500. The size on $R$ is equal to 500 with probability roughly equal to 0.1, 499 with probability around 0.56, and 498 with probability near 0.26. Thus with some degree of confidence, we can claim that for this rate-1/2 irregular code ensemble the random technique given in Algorithm \ref{alg:findR} yields a puncturing pattern with size nearly equal (and equal in some cases) to $N - k$.
\end{example}

As a direct result, a puncturing pattern generated for the irregular code of the example has a unique property. Namely, that for some patterns $D_{MP} = D_{ML}$.
\begin{lemma}\label{lem:degsFree}
Let $R_c$ denote the indices of the channel-erased bits of $p^i$, and $D_{MP}$ and $D_{ML}$ denote the degrees of freedom using MP decoding and ML decoding, respectively. If an irregular LDPC code is employed over the BEC with intentional puncturing determined by Algorithm \ref{alg:findR} in which $|R| = N - k$, then $D_{ML} = D_{MP} = |R_c|$.
\end{lemma}
\begin{IEEEproof} The ML portion of this lemma follows from Proposition \ref{prop:boundOnSizeKbar}, i.e. that the system of equations in (\ref{eq:systemML}) can resolve a maximum of $N - k$ erasures. Since $|R|=N-k$, any erasure by the channel is guaranteed to give a degree of freedom in the decoder. The MP case is the same because by Proposition \ref{prop:maxDecodMP}, the MP decoder can correct at most $N-k$ erasures. Thus any bits erased by the channel (or perhaps another set of bits of equal size) must be guessed in order to decode. Therefore, the effectiveness of the ML decoder is equal to that of the MP decoder when $|R|=N-k$.
\end{IEEEproof}

It should be noted that if the sum of systematic bits in $R+R_c$ is less than $D$, a brute-force attack on these bits might be more appealing to an attacker than decoding the entire codeword. To cover this possibility, $D$ can be thought of as the minimum between the number of systematic bits missing to the eavesdropper, and the degrees of freedom in the decoder. Although, in practice the number of systematic bits removed through puncturing or erased by the channel usually exceeds the degrees of freedom in the decoder.

\subsection{Interleaving}\label{subsec:interleaving}

The role of the interleaver is to ensure that all packets must be obtained error-free for successful decoding in any and all encoded blocks. To do this, we construct a collection of $\eta$ packets to be transmitted $X = (x^1, x^2, \ldots, x^\eta)$ in the following manner. Alice defines $\alpha$ a small positive integer which is assumed to divide $n$ (not necessary but convenient for notation and analysis) such that $\eta = n/\alpha$, and the $i$th packet is formed as
\begin{eqnarray}\label{eq:packets}
  x^i &=& (x^i_1,x^i_2,\ldots,x^i_{\alpha L}) \nonumber \\ &=&(p^1_{(i-1)\alpha+1},\ldots,p^1_{i\alpha}, p^2_{(i-1)\alpha+1},\ldots,p^2_{i\alpha},\ldots, \nonumber \\
  &&p^L_{(i-1)\alpha+1},\ldots,p^L_{i\alpha}).
\end{eqnarray}
for $i = 1,2,\ldots,\eta$. In words, we form the packet $x^i$ by concatenating $\alpha$ bits from each encoded and punctured block $p^j$ for $j = 1,2,\ldots,L$. Therefore, a single erased packet causes $\alpha$ erasures in each punctured block at the decoder. Since we have designed $R$ so that any erasure of a bit in $p^j$ results in MP decoding failure, we can be assured that any erased packet will cause all $L$ blocks to fail in the MP decoder due to this interleaving. If $R$ can be designed so that $|R| = N - k$, then the same result holds for ML decoding by Lemma \ref{lem:degsFree}.
\begin{corollary}\label{lem:security}
If $|R|=N-k$ and packets are formed according to (\ref{eq:packets}), then the number of degrees of freedom in the $i$th codeword is $D^i_{ML} = D^i_{MP} = |R^i_c| = \alpha |R_p|$ for $i=1,2,\ldots,L$, where $R_p$ is a list of all erased packets. Furthermore, $D^i_{ML} = D^j_{MP} \forall i,j$.
\end{corollary}
\begin{IEEEproof}
 The first part is trivial and follows directly from Lemma \ref{lem:degsFree} and (\ref{eq:packets}). We see that $D^i_{ML} = D^j_{MP}$ because a missing packet means exactly $\alpha$ degrees of freedom in each block, irrespective of decoder choice.
\end{IEEEproof}

\section{Decoder for Legitimate Users}\label{sec:decoder}

\begin{figure}
\begin{center}
  \begin{tikzpicture}
  [node distance=0.35cm, rounded corners=2pt, boxedNode/.style={rectangle,draw,fill=black!10,thick,
  text centered, minimum size=8mm},
  boxMultiLine/.style={rectangle,draw,fill=black!10,thick,
  text centered, text width=1.3cm, minimum size=8mm},
  inner sep=1mm]
    \node [boxedNode] (Buffer) {Buffer};
    \node [boxedNode] (Deinterleaver)  [right=of Buffer]  {Deinterleaver};
    \node [boxMultiLine] (messPass)  [right=of Deinterleaver]  {Message Passing};
    \node [rectangle,draw,fill=black!10,thick,text centered, text width=1cm, minimum size=8mm] (map)  [right=of messPass]  {Map to $\mathcal{M}$};
    \draw[->] ($(Buffer.west) + (-3mm,0)$) -- node [midway, above] {$Y$} node [midway, below=12pt, text width=1.6cm, text centered] {$\eta$ packets size $\alpha L$} (Buffer);
    \draw[->] (Buffer) to (Deinterleaver);
    \draw[->] (Deinterleaver) to node [above] {$\tilde{P}$} node [below=14pt, text width=1.5cm, text centered, midway] {$L$ blocks length $n$} (messPass);
    \draw[->] (messPass) to node [above] {$\tilde{B}$} node [below=14pt, text width=1.5cm, text centered, midway] {$L$ blocks length $N$} (map);
    \draw[->] (map) to node [midway, above] {$\tilde{M}$} node [midway, below=14pt, text width=1.5cm, text centered] {$L$  blocks length $k$} ($(map.east) + (3mm,0)$); 
  \end{tikzpicture}
\end{center}
\caption{Detailed block diagram of Bob's decoder. Number and size of blocks or packets are indicated at each step.}\label{fig:decoder}
\end{figure}
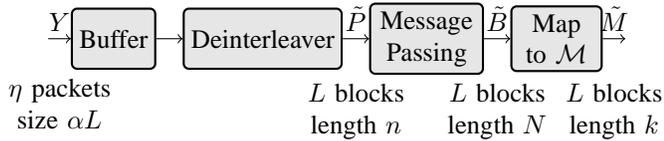

The decoder for legitimate users is simply the inverse of all encoder operations. A user can decode all data as long as every packet is received error-free. Legitimate users make use of the authenticated feedback channel to request retransmission of packets erased in the main channel during transmission. Time delay and queueing aspects of ARQ protocols are well-addressed in the literature, e.g. \cite{Konheim80} and its references. The decoding process is shown pictorially in Fig. \ref{fig:decoder}. Once all packets are obtained in $Y$, the bits are deinterleaved back into their intentionally punctured codewords $\tilde{P}$. The MP decoder is then guaranteed to decode the puncturing in linear time with the blocklength to obtain $\tilde{B}$ \cite{MoonArches}, and the inverse of the scrambling matrix is applied to the systematic decoded bits using (\ref{eq:inverseScramble}) to obtain $\tilde{M}$. Once all packets are known, this decoder guarantees that $\tilde{M} = M$.

\section{Security against Wiretappers}\label{sec:analysis}

An eavesdropper can decode the data using Bob's decoder in Fig. \ref{fig:decoder} if all packets are obtained error-free. The independence of $Q_m$ and $Q_w$, however, prevents Eve from receiving packets as a function of $\delta$ and $\epsilon$, the respective probabilities of erasures in $Q_m$ and $Q_w$. Let $R_{ef}$ be the event that a single packet is \emph{received error-free} by at least one eavesdropper after all retransmissions of the packet requested by any legitimate receiver have been filled. This section shows the blanket security effect of our encoder over nearly the entire region of possible $(\delta,\epsilon)$ pairs by completely characterizing $D$ for the system. We first show $D$ to be binomially distributed, and then provide security results for all scenarios studied as a function of $R_{ef}$. Expressions for $R_{ef}$ follow for the wiretap channel case, the broadcast scenario with $m$ intended receivers, the case with $l$ collaborating eavesdroppers, and the most general case with both $m$ legitimate receivers and $l$ collaborating eavesdroppers. For cases beyond the simple wiretap scenario, all $m$ legitimate receivers are given access to the feedback channel, and all $l$ eavesdroppers are restricted to passive status through authentication on the channel. Retransmissions in the ARQ protocol are executed only after requests are received from all legitimate parties.

Since proper design of the encoder was shown to cause $D$ to have the same realization for every codeword and be independent of the decoder in Corollary \ref{lem:security}, we understand $D$ to represent the degrees of freedom in every codeword assuming either the ML or MP decoder for the rest of the paper.

\subsection{General Security Theorems}

\begin{lemma}\label{lem:DisBinomial}
 The random variable $D$ which governs the number of degrees of freedom in a received codeword is a scaled binomial random variable. Thus, for $1 \leq \beta \leq \alpha \eta$,
 \begin{equation}\label{eq:thm2}
   \Pr(D\geq\beta) = 1 - \sum_{i=0}^{\lceil\beta/\alpha\rceil - 1}{\eta \choose i}(1-\Pr(R_{ef}))^i \Pr(R_{ef})^{\eta-i}.
 \end{equation}
\end{lemma}
\begin{IEEEproof}
 By definition, packets are erased for eavesdroppers with probability $(1-\Pr(R_{ef}))$. Since there are $\eta$ independent Bernoulli trials, each identically distributed, the sum of erased packets $|R_p|$ is a binomial random variable with parameters $\eta$ and $(1-\Pr(R_{ef}))$ \cite{Grimmett_ProbBook}. Then, by Corollary \ref{lem:degsFree}, $D = \alpha |R_p|$ where $\alpha$ bits from every codeword are sorted into each packet. Thus, $D$ is a scaled binomial random variable; specifically $D\sim \textnormal{Bin}(\eta,1-\Pr(R_{ef}))\alpha$. Since $D = \alpha|R_p|$, then $D\geq\beta$ implies that $\alpha|R_p|\geq\beta$. Clearly, this requires that $|R_p|\geq \lceil\beta/\alpha\rceil$. The result in (\ref{eq:thm2}) follows directly.
\end{IEEEproof}

The expected value is therefore known due to the binomial structure of $D$. We also prove an important property in regards to $E[D]$.
\begin{theorem}\label{thm:expectedValD}
 If $|R|=N-k$ in the encoder, then $k/n = 1$, and $E[D] = H(X|Z) = (1-\Pr(R_{ef}))n$.
\end{theorem}
\begin{IEEEproof}
 Since $|R|=N-k$, then $n=|Q|=N-|R|=k$. Let us consider the model for a single codeword ($L=1$). We can then assume $\eta$ independent uses of a PEC with packets of length $\alpha$. Let $X$ be the input to the channel, and $Z$ the output, where $\alpha$ bits are erased with probability $(1-\Pr(R_{ef}))$ or received error-free with probability $\Pr(R_{ef})$ with each channel use. The input distribution on $\alpha$ bits is uniform because the input distribution on $M$ is uniform, and the encoding function of rate one forms a bijection on $k$ bits. Thus, $H(X) = \alpha$. Clearly $H(Z|X) = H(1-\Pr(R_{ef}))$, and $H(Z) = H(1-\Pr(R_{ef})) + \Pr(R_{ef})\alpha$ (see \cite{Cover}, pg. 188). Then,
 \begin{eqnarray}
  H(X|Z) &=& H(Z|X) - H(Z) + H(X)\\
  &=& \alpha(1-\Pr(R_{ef})).
 \end{eqnarray}
 Therefore, with $\eta$ independent uses of the channel (one for each packet), $H(X|Z) = (1-\Pr(R_{ef}))\eta\alpha = (1-\Pr(R_{ef}))n$. Since the mean of a binomial random variable is the product of its two parameters, $E[D/\alpha] = (1-\Pr(R_{ef}))\eta$, and therefore
 \begin{eqnarray}
  E[D] = (1-\Pr(R_{ef}))\eta\alpha = (1-\Pr(R_{ef}))n.
 \end{eqnarray}
\end{IEEEproof}
Thus we see that $E[D]$ is equal to the information-theoretic value of equivocation when the puncturing is accomplished so that $|R|=N-k$. Therefore, \emph{perfect} secrecy is obtained when $E[D] = k$. Of course, this occurs when $\Pr(R_{ef})=0$, which implies that the eavesdropper obtains zero packets. Thus, this scheme cannot achieve perfect secrecy. However, it can be shown using the achievable rates in \cite{Wyner84} that $E[D]$ approaches the maximum achievable equivocation for $k/n = 1$. These results now require expressions for $\Pr(R_{ef})$ to complete the security characterization in $D$.

\subsection{One Receiver and One Wiretapper}

The simplest case matches the setup given in Fig. \ref{fig:wiretapChannel}, and was originally proved in \cite{Harrison10_ITW}.
\begin{lemma}[Harrison, et. al. \cite{Harrison10_ITW}]\label{lem:wiretap1}
 In the wiretap channel scenario with feedback, the probability that Eve obtains a single transmitted packet is given as
 \begin{equation}\label{eq:thm1}
   \Pr(R_{ef}) = \frac{1-\epsilon}{1-\epsilon \delta}.
 \end{equation}
\end{lemma}

Intuition of security for the wiretap channel in terms of $D$ can be gained by using the expression for $\Pr(R_{ef})$ in (\ref{eq:thm1}) to plot (\ref{eq:thm2}) for different values of $\beta$, $\alpha$, and $\eta$. Fig. \ref{fig:oneUser} shows $\Pr(D\geq 1)$ for  $\eta = 100$. Note that when $\beta=1$, $\alpha$ is not required to evaluate (\ref{eq:thm2}). This case is provided to show the plateau and falloff regions in the $(\delta, \epsilon)$ grid for $\Pr(D\geq\beta)$. Throughout the plateau region, stopping sets occur in the MP decoder and the ML decoder has linearly dependent columns in $H_{\bar{\mathcal{K}}}$ with probability very close to one. The results of Lemmas \ref{lem:DisBinomial} and \ref{lem:wiretap1} give $\Pr(D\geq1) = 1-\left(\frac{1-\epsilon}{1-\epsilon\delta}\right)^\eta$, which can be examined in the limit as $\eta\rightarrow \infty$. It is immediate that except for when $\delta = 1$ or $\epsilon = 0$, $\Pr(D\geq1)$ goes to one for all $(\delta,\epsilon)$ pairs as $\eta$ gets large. From Theorem \ref{thm:expectedValD}, if $|R|=N-k$, then $\eta = \frac{n}{\alpha} = \frac{k}{\alpha}$. Clearly $\eta$ grows with $k$; therefore, the probability of security approaches one as $k$ gets large. Since large $k$ necessitates large $n$ and $N$, the same holds true for these blocklength parameters. Codes with blocklength $N = 10,000$ are deemed practical by today's standards. For $\alpha = 1$ and for a carefully chosen $R$ with size roughly 5000, then $\eta \approx 5000$. This case is shown in Fig. \ref{fig:oneUserN5000}, where as expected, all nontrivial $(\delta,\epsilon)$ pairs show $\Pr(D\geq1)\approx 1$.

\begin{figure}
\begin{center}
  \begin{lpic}{singleUserBeta1(0.6,0.6)}
    \lbl[t]{42,8;$\epsilon$}
    \lbl[t]{100,8;$\delta$}
    \lbl[t]{75,110;$\Pr(D\geq1)$ with  $\eta = 100$}
  \end{lpic}
\end{center}
  \caption{$\Pr(D\geq1)$ when $\eta = 100$, as a function of the respective erasure probabilities in $Q_m$ and $Q_w$, $\delta$ and $\epsilon$.} \label{fig:oneUser}
\end{figure}

\begin{figure}
\begin{center}
  \begin{lpic}{singleUserN5000(0.6,0.6)}
    \lbl[t]{42,8;$\epsilon$}
    \lbl[t]{100,8;$\delta$}
    \lbl[t]{75,110;$\Pr(D\geq1)$ with  $\eta = 5000$}
  \end{lpic}
\end{center}
  \caption{$\Pr(D\geq1)$ when $\eta = 5000$, as a function of the respective erasure probabilities in $Q_m$ and $Q_w$, $\delta$ and $\epsilon$.} \label{fig:oneUserN5000}
\end{figure}

But of course, a single degree of freedom is easily guessed in an attack. Let us examine the effects on security when $\beta$ takes on a larger value. This perspective is provided in Fig. \ref{fig:oneUserN5000Beta50}, where $\eta = 5000$ and $\beta = 50$ with $\alpha = 1$. As can be seen in the figure, there exists a cutoff region, where $(\delta,\epsilon)$ pairs within the plateau region will experience at least $\beta$ degrees of freedom with probability very close to one, while pairs outside the region will have $D<\beta$ with probability close to one. Owing to the severity of the cutoff, the threshold can be approximated by setting $\Pr(D\geq\beta) = 0.5$ in (\ref{eq:thm2}), and deriving a function of $\delta$ and $\epsilon$. This technique provides a unique threshold for each specific set of values for $\beta$, $\alpha$, and $\eta$.

\begin{figure}
\begin{center}
  \begin{lpic}{singleUserN5000Beta50(0.6,0.6)}
    \lbl[t]{42,8;$\epsilon$}
    \lbl[t]{100,8;$\delta$}
    \lbl[t]{75,110;$\Pr(D\geq50)$ with $\alpha = 1$ and $\eta = 5000$}
  \end{lpic}
\end{center}
  \caption{$\Pr(D\geq50)$ when $\alpha = 1$ and $\eta = 5000$, as a function of the respective erasure probabilities in $Q_m$ and $Q_w$, $\delta$ and $\epsilon$.} \label{fig:oneUserN5000Beta50}
\end{figure}

Finally, let us inspect the $E[D]$ according to Theorem \ref{thm:expectedValD} for this case.
\begin{equation}\label{eq:ED}
  E[D] = \frac{\epsilon(1-\delta)}{1-\epsilon\delta}\eta\alpha = \frac{\epsilon(1-\delta)}{1-\epsilon\delta}n.
\end{equation}
This function grows linearly with $n$ which is equal to $k$ when $|R|=N-k$. Thus, to drive $D$ to a large number in practice, we simply must use a larger dimension in the encoder. Note that in the expectation the choice of $\alpha$ does not affect security; although, $\alpha = 1$ allows $\eta$ to be as large as possible, which provides more confidence that $D \approx E[D]$ by the law of large numbers (\cite{Grimmett_ProbBook}, pg. 193).

\subsection{Multiple Intended Receivers}
In this section, we move past the single user case, and address the more general broadcast channel originally presented in \cite{Cover72}. There is also a single eavesdropper with probability of an erased packet equal to $\epsilon$ as before. This case allows us to understand the repercussions on security of having more than one user for which we allow feedback requests. We can characterize security using Lemma \ref{lem:DisBinomial} and Theorem \ref{thm:expectedValD} in the $m$ user case by finding an expression for $\Pr(R_{ef})$. Recall that $R_{ef}$ is the event that Eve receives a single transmitted packet as before. Let each user have an independent PEC with probability of erasure in the $i$th user's channel as $\delta_i$ for $i = 1,2,\ldots,m$. The following lemma is necessary to obtain $\Pr(R_{ef})$.
\begin{lemma}\label{lem:maxMusersGeomRVs}
 If $Q_1, Q_2, \ldots ,Q_m$ are independent geometrically distributed random variables with success parameters $\lambda_1,\lambda_2,\ldots,\lambda_m$, and $T_m = \max(Q_1,Q_2,\ldots,Q_m)$, then the probability mass function on $T_m$ is given as
 \begin{equation}
   f_m(t) = \prod_{i=1}^m(1-(1-\lambda_i)^t) - \prod_{j=1}^m(1-(1-\lambda_i)^{t-1}).
 \end{equation}
\end{lemma}
\begin{IEEEproof}
 The proof is omitted for the sake of brevity, but follows from an inductive assumption on $m$.
\end{IEEEproof}

Armed with this lemma, we can obtain $\Pr(R_{ef})$ for the broadcast channel case.
\begin{lemma}\label{thm:3}
 Using the broadcast channel with $m$ independent legitimate receivers and an eavesdropper
 \begin{eqnarray}\label{eq:thm3}
  \Pr(R_{ef}) &=& \sum_{i=1}^m\left(\frac{1-\epsilon}{1-\epsilon\delta_i}\right) - \sum_{i<j}\left(\frac{1-\epsilon}{1-\epsilon\delta_i\delta_j}\right) + \nonumber \\ && \sum_{i<j<k}\left(\frac{1-\epsilon}{1-\epsilon\delta_i\delta_j\delta_k}\right)  - \cdots + \nonumber \\ && (-1)^{m+1}\left(\frac{1-\epsilon}{1-\prod_{i=1}^m\delta_i}\right)\nonumber
 \end{eqnarray}
 where the notation $i<j$ means the summation traverses over all pairs $(i,j)$ such that $i,j\in\{1,2,\ldots,m\}$ and $i<j$, and similarly for $i<j<k$, etc.
\end{lemma}
\begin{IEEEproof}
Note that if the $i$th user requests a single packet until it is received, and in each transmission it is received with probability $\delta_i$, then the total number of times the user must request the packet is governed by a geometric random variable with success parameter $1-\delta_i$ \cite{Grimmett_ProbBook}. Define $W_1, W_2, \ldots, W_m$ as the geometric random variables governing the total number of transmissions necessary for users $1,2,\ldots,m$, respectively, to obtain the packet error-free. Then, let $W = \max(W_1,W_2,\ldots,W_m)$. $W$ governs the total number of transmissions necessary for all legitimate parties to receive the packet.

By Lemma \ref{lem:maxMusersGeomRVs}, we know that
\begin{equation}
 \Pr(W=w) = \prod_{i=1}^m(1-\delta_i^w) - \prod_{j=1}^m(1-\delta_i^{w-1})
\end{equation}
because the success parameter for $W_i$ is $1-\delta_i$ for $i=1,2,\ldots,m$. Finally, we point out that
\begin{equation}
 \prod_{i=1}^m(1-\delta_i) = 1 - \sum_{i=1}^m\delta_i + \sum_{i<j}\delta_i\delta_j - \sum_{i<j<k}\delta_i\delta_j\delta_k + \cdots (-1)^m\prod_{i=1}^m\delta_i
\end{equation}
which implies that
\begin{eqnarray}
 \Pr(W=w) &=& \left(1 - \sum_{i=1}^m\delta_i^w + \sum_{i<j}(\delta_i\delta_j)^w - \cdots + \right.\nonumber \\ &&\left.(-1)^{m}(\prod_{i=1}^m\delta_i)^w\right) + \nonumber \\
 && \left(-1 + \sum_{i=1}^m\delta_i^{w-1} - \sum_{i<j}(\delta_i\delta_j)^{w-1} + \cdots + \right.\nonumber \\ &&\left. (-1)^{m+1}(\prod_{i=1}^m\delta_i)^{w-1}\right) \nonumber \\
 &=& \sum_{i=1}^m\delta_i^{w-1}(1-\delta_i) - \sum_{i<j}(\delta_i\delta_j)^{w-1}(1-\delta_i\delta_j)  \nonumber \\ && + \cdots + (-1)^{m+1}(\prod_{i=1}^m\delta_i)^{w-1}(1-\prod_{i=1}^m\delta_i). \nonumber
\end{eqnarray}

With these pieces in place, we commence proving the lemma.
\begin{eqnarray}\label{eq:longListEqns}
 \Pr(R_{ef}) &=& \sum_{w=1}^\infty\Pr(R_{ef}|W=w)\Pr(W=w) \nonumber \\
 &=& \sum_{w=1}^\infty(1-\epsilon^w)\left(\prod_{i=1}^m(1-\delta_i^w) - \prod_{j=1}^m(1-\delta_i^{w-1})\right) \nonumber \\
 &=& \sum_{w=1}^\infty(1-\epsilon^w)\left(\sum_{i=1}^m\delta_i^{w-1}(1-\delta_i) - \right. \nonumber \\ & & \sum_{i<j}(\delta_i\delta_j)^{w-1}(1-\delta_i\delta_j) + \cdots \nonumber \\ & & + \left. (-1)^{m+1}(\prod_{i=1}^m\delta_i)^{w-1}(1-\prod_{i=1}^m\delta_i)\right) \nonumber \\
 &=& \sum_{i=1}^m\frac{1-\delta_i}{\delta_i}\sum_{w=1}^\infty(1-\epsilon^w)\delta_i^w - \nonumber \\ & & \sum_{i<j}\frac{1-\delta_i\delta_j}{\delta_i\delta_j}\sum_{w=1}^\infty(1-\epsilon^w)(\delta_i\delta_j)^w + \cdots \nonumber \\ & & + (-1)^{m+1}\frac{1-\prod_{i=1}^m\delta_i}{\prod_{i=1}^m\delta_i}\sum_{w=1}^\infty(1-\epsilon^w)(\prod_{i=1}^m\delta_i)^w \nonumber \\
 &=& \sum_{i=1}^m\frac{1-\delta_i}{\delta_i}\left(\sum_{w=0}^\infty\delta_i^w - \sum_{w=0}^\infty(\epsilon\delta_i)^w\right) - \nonumber \\ & & \sum_{i<j}^m\frac{1-\delta_i\delta_j}{\delta_i\delta_j}\left(\sum_{w=0}^\infty(\delta_i\delta_j)^w - \sum_{w=0}^\infty(\epsilon\delta_i\delta_j)^w\right)  \nonumber \\ & & + \cdots + (-1)^{m+1}\frac{1-\prod_{i=1}^m\delta_i}{\prod_{i=1}^m\delta_i}\times \nonumber \\ & & \left(\sum_{w=0}^\infty(\prod_{i=1}^m\delta_i)^w - \sum_{w=0}^\infty(\epsilon\prod_{i=1}^m\delta_i)^w\right) \nonumber\\
 &=& \sum_{i=1}^m\left(\frac{1-\epsilon}{1-\epsilon\delta_i}\right) - \sum_{i<j}^m\left(\frac{1-\epsilon}{1-\epsilon\delta_i\delta_j}\right) + \cdots + \nonumber \\ & & (-1)^{m+1}\left(\frac{1-\epsilon}{1-\prod_{i=1}^m\delta_i}\right).
\end{eqnarray}
\end{IEEEproof}

\subsection{Collaborating Eavesdroppers}
In this section we consider the case with $l$ eavesdroppers working together in order to obtain the cryptogram $M$, each with a possibly unique probability of packet erasure $\epsilon_1, \epsilon_2, \ldots, \epsilon_l$. All are assumed to obtain packets through independent PECs. It is simpler to first consider a single legitimate user Bob with probability of packet erasure $\delta$. Then the general result which assumes $m$ friendly parties with $l$ collaborating eavesdroppers comes easily.
\begin{lemma}\label{thm:5}
 For $l$ eavesdroppers and a single legitimate receiver,
 \begin{equation}
  \Pr(R_{ef}) = \frac{1 - \prod_{i=1}^l\epsilon_i}{1-\delta\prod_{i=1}^l\epsilon_i}.
 \end{equation}
\end{lemma}
\begin{IEEEproof}
 The proof is straightforward if we note that collaborating eavesdroppers receive a single sent packet if at least one of them obtains the packet error-free. Let $W$ be a geometric random variable with success parameter $1-\delta$. This governs the number of transmissions for each packet. Therefore,
 \begin{eqnarray}\label{eq:Ref1userLeaves}
  \Pr(R_{ef}) &=& \sum_{w=1}^\infty\Pr(R_{ef}|W=w)\Pr(W=w)\nonumber \\
  &=& \sum_{w=1}^\infty(1-(\prod_{i=1}^l\epsilon_i)^w)(1-\delta)\delta^{w-1} \nonumber \\
  &=&\frac{1-\delta}{\delta}\left(\sum_{w=0}^\infty\delta^w - (\delta\prod_{i=1}^l\epsilon_i)^w\right) \nonumber \\
  &=& \frac{1-\prod_{i=1}^l\epsilon_i}{1-\delta\prod_{i=1}^l\epsilon_i}.
 \end{eqnarray}
\end{IEEEproof}

This answer provides an easy bridge to an extremely general result.
\begin{corollary}\label{thm:6}
For the scenario with $m$ intended parties and $l$ eavesdroppers with similar notation as before,
 \begin{eqnarray}
  \Pr(R_{ef}) &=& (1-\epsilon')\left(\sum_{i=1}^m\frac{1}{1-\epsilon'\delta_i} - \sum_{i<j}\frac{1}{1-\epsilon'\delta_i\delta_j} + \cdots + \nonumber \right.\\ & & \left. (-1)^{m+1}\frac{1}{1-\epsilon'\prod_{i=1}^m\delta_i}\right),
 \end{eqnarray}
 where $\epsilon' = \prod_{i=1}^l\epsilon_i$.
\end{corollary}
\begin{IEEEproof}
 This proof is not included for the sake of brevity, but is nearly identical to the proof of Lemma \ref{thm:3} with slight alterations as indicated by the proof of Lemma \ref{thm:5} to allow for multiple eavesdroppers.
\end{IEEEproof}

\section{Cryptographic Security Enhancements}\label{sec:crypto}
The probabilistic security analysis in Section \ref{sec:analysis} assumes that attacks on the cryptography become more difficult or completely infeasible as $D$ gets large. It remains to show the effect of the coding scheme on attacks of the cryptography. As an example, fast correlation attacks on stream ciphers are known to be possible, even if the cryptogram is error-prone. It was noted in \cite{Harrison09_ICC,Harrison09_ISIT,Harrison09_ITW} that specific attacks from \cite{Meier89} were made more difficult, and in some cases impossible due to error rates in the cryptogram beyond a certain threshold. Certainly as bit error rates approach 0.5 in the cryptogram, attacks of the fast-correlation variety break down completely.

Let $\hat{P}=(\hat{p}^1,\hat{p}^2,\ldots,\hat{p}^L)$ be the collection of punctured codewords obtained by Eve, where $\hat{p}^i = (\hat{p}^i_1,\hat{p}^i_2,\ldots,\hat{p}^i_n)$, and let $\hat{B}=(\hat{b}^1,\hat{b}^2,\ldots,\hat{b}^L)$ be the decoded codewords, where $\hat{b}^i = (\hat{b}^i_1,\hat{b}^i_2,\ldots,\hat{b}^i_N)$. Finally, define the implied block structure of Eve's decoder output as $\hat{M}=(\hat{m}^1,\hat{m}^2,\ldots,\hat{m}^L)$, where $\hat{m}^i=(\hat{m}^i_1,\hat{m}^i_2,\ldots,\hat{m}^i_k)$. Each channel-erased bit in $\hat{p}^i$ yields a degree of freedom in $\hat{b}^i$, and complete recovery of $\hat{b}^i$ requires that $D$ bits in $\hat{p}^i$ be guessed correctly. If a guess is incorrect, there will be at least as many errors in $\hat{b}^i$ as the minimum distance of the LDPC code. The descrambling process in (\ref{eq:inverseScramble}) magnifies any errors in $\hat{b}^i$ to an expected bit error rate of 0.5 in $\hat{m}^i$. Therefore, since all guesses are equally likely, a brute-force attack on $D$ bits must be accomplished to obtain each $\hat{m}^i$.

\begin{figure}
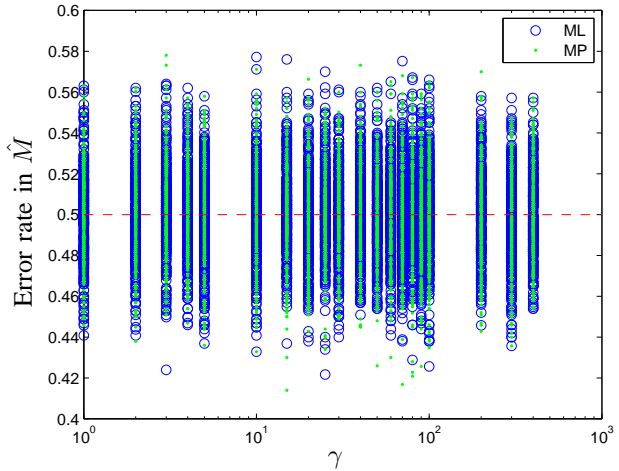

\begin{center}
  \begin{lpic}{cryptoErrorRates(0.6,0.6)}
    \lbl[t]{75,5;$\gamma$}
    \lbl[t]{75,113; Error propagation for incorrect guesses}
    \lbl[l]{5,37,90;Error rate in $\hat{M}$}
  \end{lpic}
\end{center}
  \caption{The simulated error rates in Eve's decoded cryptogram $\hat{M}$ when $\gamma$ errors are made in guessing bit values for $D$ degrees of freedom in Eve's received codewords.} \label{fig:cryptoErrorRates}
\end{figure}

Simulations of the end-to-end encoder and decoder clearly indicate the expected bit error rate in $\hat{M}$ of 0.5 for an incorrect guess. Simulations were performed using the irregular LDPC code of Example \ref{ex:punct} with $N=1000$ and $k=500$. Puncturing patterns used were such that $|R|\geq 498$ bits. $S$ was formed randomly by setting roughly half of the $k^2$ entries equal to one until such a matrix was invertible using the LU decomposition in GF(2). Let $\gamma$ be the number of bits in Eve's guess which are incorrect. We offer simulation results for $\gamma =$ 1, 2, 3, 4, 5, 10, 15, 20, 25, 30, 40, 50, 60, 70, 80, 90, 100, 200, 300, and 400 in Fig. \ref{fig:cryptoErrorRates}. Each $\gamma$ value was tested 300 times on both the MP and ML decoder, while a new puncturing pattern $R$ was generated every 10 experiments, and a new code from the ensemble was selected every 30 experiments. All tests produced error rates in between 0.414 and 0.578 in $\hat{M}$, while the mean depicted a 0.5002 bit error rate with no noticeable difference between MP and ML decoders, or between $\gamma$ values, as Fig. \ref{fig:cryptoErrorRates} indicates.

These results imply that unless $D$ bits are guessed exactly, the cryptography must be attacked with an average bit error rate of 0.5 in $\hat{M}$. We can certainly expect such an attack to fail for fast correlation attacks on stream ciphers, but the notion that any attack on a cryptosystem could absorb such error rates and still succeed is obviously shortsighted. However, since an attack could feasibly be staged using a single block of $\hat{M}$, we will only guarantee failure of the attack if every block in $\hat{M}$ is incorrect. Using similar logic, it can be said that if an attack would succeed using the error-free ciphertext $M$, then it may fail even if a single block in $\hat{M}$ is in error.
\begin{theorem}
 Define the complexity of a cryptographic attack to be $C_A$. Let $D$ be the degrees of freedom of each of $L$ blocks in $\hat{B}$. Then the expected complexity $C_{PL}$ of a successful attack on the system is bounded as
 \begin{equation}\label{eq:lastTheorem}
  2^{E[D]}(1-2^{-1/L})C_A \leq C_{PL} \leq 2^{E[D]}(2^{-1/L})C_A.
 \end{equation}
\end{theorem}
\begin{IEEEproof}
 By Corollary \ref{lem:security} each codeword in $\hat{B}$ has the same number of degrees of freedom. Thus, $E[D]$ is the average number of bits that must be guessed in each of $L$ punctured codewords in $\hat{P}$. Assume that an attacker guesses bit patterns on all codewords in $\hat{P}$ simultaneously. The correct bit patterns of the channel-erased bits in the $L$ codewords $\hat{P}$ are uniformly distributed over $2^{E[D]}$ possibilities in each block. The lower bound is formulated by the expected number of guesses until at least one of $L$ codewords is found. Model the correct bit patterns in the $L$ codewords as i.i.d. discrete uniform random variables on $\{0,1,\ldots,2^{E[D]}-1\}$, say $U_1, U_2, \ldots, U_L$. Without loss of generality, assume that an attacker begins by guessing zero for each $U_i$ and proceeds in an orderly fashion. Then, the expected number of guesses until at least one is correct is given by $E[\min(U_1,U_2,\ldots,U_L)]$. Thus, we calculate $\Pr(\min(U_1,U_2,\ldots,U_L)\geq z)= \Pr(U_1\geq z,\Pr(U_2\geq z),\ldots,U_L\geq z) = (\Pr(U_1\geq z)(\Pr(U_2\geq z)\ldots(\Pr(U_L\geq z) = $
 \begin{equation}
  \left(\frac{2^{E[D]}-z}{2^{E[D]}}\right)^L.
 \end{equation}
 Now, solve for $z$ in $\Pr(\min(U_1,U_2,\ldots,U_L)\geq z)=0.5$ for a close bound on the expectation to get the lower bound.

 The upper bound is calculated similarly, but we assume that \emph{all} patterns must be guessed in order to guarantee success, therefore, the bound is given by finding the $z$ that solves $\Pr(\max(U_1,U_2,\ldots,U_L)<z)=0.5$.
\end{IEEEproof}
As a check on these bounds, for $L=1$ we expect $2^{E[D]-1}$ guesses on average for a successful attack. In this case, both bounds meet at $2^{E[D]-1}C_A$, as expected. Although these bounds are helpful, when $L>1$ the bounds are not as tight, and thus provide limited insight into the true increase in complexity of the attack. More than likely, an attack will require at least a certain number of consecutive blocks in $M$ to execute successfully \cite{Meier89}. Clearly a 0.5 bit error rate in any block would destroy an attack with these requirements. Therefore, the upper bound in (\ref{eq:lastTheorem}) serves as a good approximation to the expected amount of work necessary to complete the attack, with $L$ being set by the attack specifications. Thus we see, that our system appends a multiplier which is exponential in $E[D]$ to the complexity of a cryptographic attack through practical physical-layer security.

\section{Conclusions}\label{sec:conclusion}
In conclusion, we have presented the security metric of degrees of freedom $D$ in an eavesdropper's received codewords, and applied this metric to a physical-layer coding scheme to show cryptographic security enhancements due to channel coding. The coding scheme relies on the nature of independent packet erasure channels and ARQ to provide secrecy and reliability, respectively. End-to-end details of the encoder and decoder were provided. Design criteria were specified to maximize $D$ in a maximum-likelihood attack as well as a message-passing attack. This involved security performance comparisons of LDPC codes with varying degree distributions, where irregular codes were shown to outperform regular codes in maximizing $D$. The expected value of $D$ was also shown to be equal to $H(X|Z)$ in our encoder. Probabilistic security results were obtained and made general so as to apply to multiple receivers and multiple collaborative attackers. Simulation results were provided which show that unless an attacker can guess $D$ symbols in the received data correctly, the system yields a bit error rate of 0.5 in the cryptogram, thus necessitating a brute-force attack on $D$ bits for each codeword. The end result on the expected increase in attack complexity on the cryptosystem due to our scheme is a multiplier which is exponential in $E[D]$. The system was shown to provide cryptographic security enhancement, even when eavesdroppers have an advantage over legitimate receivers in signal quality.


\bibliographystyle{ieeetran}
\bibliography{references_StopSetsJournal}

\ifCLASSOPTIONpeerreview
\pagebreak
\listoffigures
\listoftables
\fi

\end{document}